\documentclass[journal]{IEEEtran}

\ifCLASSINFOpdf
\else
   \usepackage[dvips]{graphicx}
\fi
\usepackage{url}
\usepackage{algorithm}
\usepackage{algpseudocode}
\usepackage{amsthm}
\usepackage{amsmath}
\usepackage{amssymb}
\usepackage{enumerate}
\usepackage{mathtools}
\usepackage{graphicx}
\usepackage[caption=false,font=footnotesize]{subfig}
\usepackage{cite}
\usepackage{hyperref}
\hypersetup{
    colorlinks=true,
    citecolor=blue,
    linkcolor=blue,
    urlcolor=blue
}

\usepackage{graphicx}

\newtheorem{definition}{Definition}
\newtheorem{lemma}[definition]{Lemma}
\newtheorem{theorem}[definition]{Theorem}

\theoremstyle{remark}

\begin{document}

\title{Scaled-Gradient Hard Thresholding Algorithms for Sparse Signal Recovery}

\author{Nan Meng, Yun-Bin Zhao
\thanks{This work was supported by the Natural Science Foundation of China under the grants 12301400, 12471295 and 12426306, and Hetao Shenzhen-Hong Kong Science and Technology Innovation Cooperation Zone Project (HZQSWS-KCCYB-2024016). }
\thanks{Nan Meng is with the School of Mathematical Sciences,
University of Nottingham Ningbo China, Ningbo 315100, China
(e-mail: Nan.Meng@nottingham.edu.cn).}
\thanks{ Yun-Bin Zhao (Coresponding author) is with the Shenzhen International Center for
Industrial and Applied Mathematics, SRIBD, The Chinese University of Hong Kong, Shenzhen, 
China (e-mail: Yunbinzhao@cuhk.edu.cn).}}
\maketitle

\begin{abstract}
Thresholding methods provide a simple iterative framework for sparse signal recovery by combining a search direction with a thresholding operation.
Gradient-based search directions are computationally efficient because they require only first-order information for the error metric.
We incorporate componentwise gradient scaling into this framework by preserving large-magnitude components and attenuating the remaining components of the gradient of error metric.
Combining this strategy with hard thresholding yields scaled-gradient hard thresholding (SGHT) and its pursuit variant SGHTP.
Under suitable restricted isometry conditions, we establish recovery error bounds for both methods that account for measurement noise and deviations from exact sparsity.
Numerical experiments indicate that SGHTP achieves recovery performance comparable to established pursuit methods on the tested ECG signals.
\end{abstract}

\begin{IEEEkeywords}
Compressed sensing, gradient-based methods, hard thresholding, sparse signal recovery, thresholding methods.
\end{IEEEkeywords}

\IEEEpeerreviewmaketitle

\section{Introduction}

\IEEEPARstart{S}{parse} signal recovery is a fundamental task in compressed sensing.
It has wide applications in image reconstruction \cite{lustig2007sparse, chen2025pcnet}, channel estimation \cite{bajwa2010compressed}, and inverse problems \cite{daubechies2004iterative}.
A typical formulation for sparse signal recovery is
\begin{equation} \label{prob}
\min _x\frac{1}{2}\|y-A x\|_2^2 \quad s.t. \quad\|x\|_0 \leq k,
\end{equation}
where $A \in \mathbb{R}^{m \times n}$ is the measurement matrix, $y \in \mathbb{R}^m$ is the observed vector, $\|x\|_0$ counts the nonzero entries in $x$, and $k \in \mathbb{Z}$ is the  sparsity level. The objective function in (\ref{prob}) is often called the error metric for singal recovery.
A few class of numerical methods have been developed for this problem, such as convex relaxations \cite{tibshirani1996regression, chen2001atomic, donoho2001uncertainty, donoho2006stability}, greedy pursuits \cite{tropp2004greed, tropp2007omp, dai2009subspace, needell2009cosamp, donoho2012sparse, wang2016recovery, zhao2023dynamic}, and thresholding methods \cite{blumensath2008iterative, blumensath2009iterative, foucart2011hard, blumensath2010normalized, tanner2013normalized, zhao2020optimal}.
Thresholding methods are particularly attractive for sparse signal recovery, because they admit exceptionally low per-iteration complexity.

In this paper, we focus on the class of hard thresholding approaches.
The typical hard-thresholding iteration consists of two operations: computing a certain search direction and performing hard thresholding.
The first hard thresholding method might be the iterative hard thresholding (IHT) \cite{blumensath2009iterative} which adopts the steepest descent direction as the search direction. 
To enhance the performance of IHT, the so-called hard thresholding pursuit (HTP) \cite{foucart2011hard} was proposed by merging a pursuit step into IHT.
Several other algorithms have also been developed in the literature, including the ones using adaptive step-sizes and acceleration \cite{blumensath2010normalized, blumensath2012accelerated}, optimal thresholding \cite{zhao2020optimal, zhao2022natural}, and those search directions distinct from the steepest descent direction \cite{meng2020newton, meng2022partial}.

So far, almost all search directions in thresholding methods can be viewed as a certain scaling of the gradient of the error metric, that is
\begin{equation} \label{direction}
d =-D\nabla f(x),
\end{equation}
where $f(x) = \|y-A x\|_2^2$, and $D \in \mathbb{R}^{n \times n}$ is a scaling matrix.
For instance, choosing $D$ as a regularized inverse Hessian yields a Newton-type search direction \cite{meng2020newton}.
If $D$ is an identity matrix, \eqref{direction} becomes the classic steepest descent direction which has been used in most existing iterative methods for  signal recovery. In this case, the scaling matrix $D$ assigns the same weight to every entry of gradient.
However,  theoretical and numerical results indicate that the large-magnitude components of gradient are more likely to influence the support identification of sparse signals.
The existing partial gradient methods \cite{meng2022partial} adopt this idea by simply using a scaling diagonal matrix $D$ with a small number of diagonal entries being 1 and the remaining zeros.
These methods usually keep only a few large-magnitude components of gradient as a search direction and completely discard the remaining smaller ones in magnitude.

However, the size of components of gradients may have unremarkable differences and might be very close to each other.
In these cases, completely abandoning large number of gradient components to construct a search direction may cause a big loss of useful gradient information, thereby slowing the convergence of the underlying algorithms. 
Thus it makes sense to propose a more reasonable scaling diagonal matrix $D$ that assigns relatively large weights to large-magnitude components of gradient and relatively small weights to small-magnitude components, but not completely discard the small-magnitude ones as in existing partial gradient methods.
This study is devoted to such a scaling strategy and the development of an iterative algorithm based on it.

The main contributions of this work are summarized as follows.
i) We introduce the scaled-gradient direction, and develop scaled-gradient hard thresholding (SGHT) together with its pursuit variant called SGHTP.
ii) We establish RIP-based convergence guarantees for both algorithms.
iii) We demonstrate some numerical results about the effects of the scaling parameters on support identification and signal recovery, and we compare the proposed methods with representative methods for sparse signal reconstruction.

\section{Algorithms}
For the error metric $f(x)=\frac{1}{2}\|y-Ax\|_2^2$, its gradient is given by $\nabla f(x)=A^\top(Ax-y)$.
The standard gradient-based hard-thresholding iteration is
\[
x^{(p+1)}=\mathcal{H}_k (x^{(p)} -\lambda \nabla f(x^{(p)})),
\]
where $\lambda>0$ is the step size and $\mathcal{H}_k$ retains the $k$ largest-magnitude entries of its argument, setting the others to zero.

\subsection{Search direction}
Given an integer $q\in\{1,\ldots,n\}$, let $\Omega^{(p)}$ be the index set of the $q$ largest-magnitude entries of $\nabla f(x^{(p)})$.
We define
\begin{equation} \label{com-gradient}
d^{(p)}=-Z^{(p)} \nabla f(x^{(p)}),
\end{equation}
where $Z^{(p)}=\operatorname{diag} (z_1^{(p)},\ldots,z_n^{(p)})$ with
\[
z_i^{(p)}=
\begin{cases}
1,      & i\in\Omega^{(p)},\\
\gamma, & i\notin\Omega^{(p)},
\end{cases}
\qquad \gamma\in[0,1].
\]
The scaling weight $Z^{(p)}$ preserves the $q$ largest-magnitude components of $  \nabla f(x^{(p)})$ and attenuates its remaining components by $\gamma$. 
Since $Z^{(p)}$ is positive semidefinite, we have
\[
\begin{aligned}
\nabla f(x^{(p)})^{\top} d^{(p)} & =-\nabla f(x^{(p)})^{\top} Z^{(p)} \nabla f(x^{(p)}) \leq 0 .
\end{aligned}
\]
Hence, $d^{(p)}$ is a strictly descent direction of $ f(x)$ at $x^{(p)}$  whenever $ d^{(p)} \neq 0$.

\subsection{SGHT and SGHTP Algorithms}
We use notation $[N]:=\{1,\ldots,n\}$ and $\operatorname{supp}(v):=\{i\in[N]:v_i\ne0\}$. Combining the search direction in \eqref{com-gradient} with hard thresholding yields SGHT and its pursuit variant SGHTP, summarized in Algorithm~\ref{algorithm}.
When $\gamma=1$, SGHT and SGHTP reduce to IHT \cite{blumensath2009iterative} and HTP \cite{foucart2011hard}, respectively. These reductions also hold when $q=n$.
When $\gamma=0$, the search direction becomes $-\mathcal H_q(\nabla f(x^{(p)}))$, as used in partial gradient optimal thresholding methods \cite{meng2022partial}.

With zero initialization, the same positive step size, and consistent rules for tie-breaking and exact least-squares selection, SGHTP coincides with HTP whenever $q\ge k$. We thus focus on $q<k$, which is necessary for the scaling to alter the pursuit iteration.

\begin{algorithm}[t]
\caption{SGHT ($b=0$) and SGHTP ($b=1$)}
\label{alg:sght-sghtp}
\begin{algorithmic}[1]
\Require $A$, $y$, $k$, $q$, $\lambda>0$,
         $\gamma\in[0,1]$, $b\in\{0,1\}$, $P_{\max}$
\State $x^{(0)}=\mathbf{0}$
\For{$p=0,\ldots,P_{\max}-1$}
    \State $\Omega^p=\operatorname{supp}\bigl(
           \mathcal{H}_q(A^\top(y-Ax^{(p)}))\bigr)$
    \State $u^{(p)}=x^{(p)}+\lambda Z^p A^\top
           \bigl(y-Ax^{(p)}\bigr)$
    \label{alg1-u}
    \State $x^{(p+1)}=\mathcal{H}_k(u^{(p)})$
    \If{$b=1$}
        \State $T^p=\operatorname{supp}(x^{(p+1)})$
        \State $x^{(p+1)}=
        \arg\min_{\operatorname{supp}(z)\subseteq T^p}
        \lVert y-Az\rVert_2^2$
    \EndIf
\EndFor
\State \Return $x^{(P_{\max})}$
\end{algorithmic} \label{algorithm}
\end{algorithm}

\section{Theoretical Analysis}
We first recall the definition of the restricted isometry constant and then state two auxiliary results used in the subsequent analysis.

\begin{definition} \cite{candes2005decoding}
For $A \in \mathbb{R}^{m \times n}$, the restricted isometry constant (RIC) of order $k$, denoted by $\delta_k$, is the smallest $\delta \geq 0$ such that
$
(1-\delta)\|x\|_2^2 \leq\|A x\|_2^2 \leq(1+\delta)\|x\|_2^2
$ 
holds for every $k$-sparse vector $x \in \mathbb{R}^n$.
The matrix $A$ satisfies the restricted isometry property (RIP) of order $k$ if $\delta_k<1$.
\end{definition}

For $\Lambda\subseteq[N]$, let $v_\Lambda$ denote the vector obtained from $v$ by setting its entries outside $\Lambda$ to zero.

\begin{lemma} \cite{foucart2013mathematical} \label{lemma-rip}
Let $A \in \mathbb{R}^{m \times n}$ satisfy the RIP of order $t$ with RIC $\delta_t$. 
For any vector $v \in \mathbb{R}^n$, $u \in \mathbb{R}^m$, and any index set $\Lambda \subseteq[N]$,\begin{enumerate}[i)]
\item $\|((I-A^T A) v)_{\Lambda}\|_2 \leq \delta_t\|v\|_2$, if $|\Lambda \cup \operatorname{supp}(v)| \leq t$;
\item $\|(A^T u)_{\Lambda}\|_2 \leq \sqrt{1+\delta_t}\|u\|_2$, if $|\Lambda| \le t$.
\end{enumerate}
\end{lemma}

\begin{lemma} \cite{zhao2023improved} \label{lemma-golden}
Let $z \in \mathbb{R}^n$, and let $v \in \mathbb{R}^n$ be $k$-sparse (i.e., $\|v\|_0 \leq k$). Then
$
\|v-\mathcal{H}_k(z)\|_2 \leq \varphi\|(v-z)_{\Lambda \cup S^*}\|_2,
$
where $\varphi=\frac{\sqrt{5}+1}{2}$, $\Lambda=\operatorname{supp}(v)$, and $S^*=\operatorname{supp}\left(\mathcal{H}_k(z)\right)$.
\end{lemma}

\subsection{Recovery Guarantees for SGHT}

Let $S$ be the index set for the $k$ largest-magnitude entries of $x$, and hence $x_S$ is its best $k$-term approximation.
For measurements $y=Ax+\eta$ with noise $\eta$, write $y=Ax_S+\eta'$, where $\eta':=Ax_{\overline S}+\eta$ and $\overline S:=[N]\setminus S$.

\begin{theorem}
Let $A \in \mathbb{R}^{m \times n}$ satisfy
\begin{equation} \label{thm1-ric}
\delta_{3 k}<\frac{(1+\gamma) \varphi^{-1}-(1-\gamma)}{2},
\end{equation}
where $\varphi = \frac{\sqrt{5}+1}{2}$.
Choose $\lambda$ such that
\begin{equation} \label{thm1-lam}
\frac{1-\varphi^{-1}}{\gamma-\delta_{3 k}}<\lambda<\frac{\varphi}{1+\delta_{3 k}} .
\end{equation}
Then, for every $p \geq 0$, we have
\[
\|x^{(p+1)}-x_S\|_2 \leq \rho_1\|x^{(p)}-x_S\|_2+\tau_1\|\eta^{\prime}\|_2,
\]
where
\begin{equation} \label{thm1-tau}
\tau_1:=\varphi \lambda \sqrt{1+\delta_{2 k}}
\end{equation}
and 
\begin{equation} \label{thm1-rho}
\rho_1:= \begin{cases}\varphi[1-\lambda(\gamma-\delta_{3 k})], & \textrm{ if } 0<\lambda \leq \frac{2}{1+\gamma} \\ \varphi[\lambda(1+\delta_{3 k})-1], & \textrm{ if }  \lambda>\frac{2}{1+\gamma}\end{cases}
\end{equation}
which is smaller than 1. 

\end{theorem}

\begin{proof}
Applying Lemma \ref{lemma-golden} with $v=x_S$ and $z=u^{(p)}$ gives
\[
\|x^{(p+1)} - x_{S}\|_2 \le \varphi \|(u^{(p)} - x_{S})_{S^{p+1} \cup S}\|_2,
\]
where $S^{p+1}=\operatorname{supp}(x^{(p+1)})=\operatorname{supp}(\mathcal{H}_k(u^{(p)}))$.
From \eqref{alg1-u} and $y = Ax_S + \eta'$, we have
\[
\begin{aligned}
u^{(p)}-x_S & =x^{(p)}-x_S+\lambda Z^p A^{\top}(A x_S+\eta^{\prime}-A x^{(p)}) \\
&=(I-\lambda Z^p)(x^{(p)}-x_S) +\lambda Z^p A^{\top} \eta^{\prime} \\
&~~~+\lambda Z^p(I-A^{\top} A)(x^{(p)}-x_S) .
\end{aligned}
\]
Since $Z^{(p)}$ is diagonal with entries in $\{1, \gamma\}$,
\[
\|I-\lambda Z^{(p)}\|_2 \leq \Delta_{\lambda, \gamma}, ~~ \Delta_{\lambda, \gamma}:=\max \{|1-\lambda|,|1-\gamma \lambda|\}.
\]
Therefore, we have
\begin{equation} \label{thm1-1}
\|[(I-\lambda Z^{(p)})(x^{(p)}-x_S)]_{S^{p+1} \cup S}\|_2 \leq \Delta_{\lambda, \gamma}\|x^{(p)}-x_S\|_2 .
\end{equation}
Moreover, since the diagonal entries of $Z^{(p)}$ lie in $[0,1]$,
\begin{equation} \label{thm1-2}
\|(Z^p v)_{\Lambda}\|_2 \leq \|v_{\Lambda}\|_2.
\end{equation}
for any $v \in \mathbb{R}^n$ and $\Lambda \subseteq[N]$. 
It is easy to see that
\[
|(S^{p+1} \cup S) \cup \operatorname{supp}(x^{(p)}-x_S)| \leq 3 k,
\]
as $x^{(p)}, x^{(p+1)}$, and $x_S$ are all $k$-sparse.
Hence, by \eqref{thm1-2} and Lemma \ref{lemma-rip} (i), we have
\begin{align}
&\|[Z^p(I-A^{\top} A)(x^{(p)}-x_S)]_{S^{p+1} \cup S}\|_2 \nonumber \\
&\le \|[(I-A^{\top} A)(x^{(p)}-x_S)]_{S^{p+1} \cup S}\|_2 
\le \delta_{3 k}\|x^{(p)}-x_S\|_2. \label{thm1-4}
\end{align}
As $|S^{p+1} \cup S| \leq 2 k$, by \eqref{thm1-2} and Lemma \ref{lemma-rip} (ii), we obtain
\begin{align} 
\|(Z^p A^{\top} \eta^{\prime})_{S^{p+1} \cup S}\|_2 \le \sqrt{1+\delta_{2 k}}\|\eta^{\prime}\|_2 , \label{thm1-5}
\end{align}
Combining \eqref{thm1-1}, \eqref{thm1-4}, and \eqref{thm1-5} gives
\[
\begin{aligned}
\|(u^{(p)} - x_{S})_{S^{p+1} \cup S}\|_2 \le & (\Delta_{\lambda, \gamma}+\lambda \delta_{3 k})\|x^{(p)}-x_S\|_2 \\
&+\lambda \sqrt{1+\delta_{2 k}}\|\eta^{\prime}\|_2 .
\end{aligned}
\]
Consequently, we have
\[
\|x^{(p+1)}-x_S\|_2 \leq \varphi(\Delta_{\lambda, \gamma}+\lambda \delta_{3 k}) \|x^{(p)}-x_S\|_2+\tau_1 \|\eta^{\prime}\|_2,
\]
where $\tau_1$ is given by \eqref{thm1-tau}.
For $\lambda>0$ and $\gamma \in[0,1]$, it is easy to verify that
\[
\Delta_{\lambda, \gamma}= \begin{cases}1-\gamma \lambda, & 0<\lambda \leq \frac{2}{1+\gamma}, \\ \lambda-1, & \lambda > \frac{2}{1+\gamma}. \end{cases}
\]
Thus, the coefficient of $\|x^{(p)}-x_S\|_2$ equals $\rho_1$ in \eqref{thm1-rho}.

It remains to verify that $\rho_1<1$.
Condition~\eqref{thm1-ric} implies $\gamma>\delta_{3k}$ and
\[
\frac{1-\varphi^{-1}}{\gamma-\delta_{3k}}
<
\frac{2}{1+\gamma}
<
\frac{\varphi}{1+\delta_{3k}},
\]
so the step-size interval in \eqref{thm1-lam} is nonempty.
For $0<\lambda\le2/(1+\gamma)$, the lower bound in \eqref{thm1-lam} gives
$ 
\rho_1=\varphi[1-\lambda(\gamma-\delta_{3k})]<1.
$ 
For $\lambda>2/(1+\gamma)$, the upper bound in \eqref{thm1-lam} gives
$
\rho_1=\varphi[\lambda(1+\delta_{3k})-1]<1.
$ 
Hence, every step size satisfying \eqref{thm1-lam} ensures $\rho_1<1$. (It is worth noting that for $\gamma=1$ and $\lambda=1$, condition \eqref{thm1-ric} reduces to $\delta_{3k}<\varphi^{-1}$, recovering the IHT guarantee in \cite{zhao2023improved}.)
\end{proof}

\subsection{Theoretical results for SGHTP}
The following lemma bounds the error introduced by the pursuit step in SGHTP.

\begin{lemma} \cite{zhao2020optimal} \label{lemma-pursuit}
Let $A \in \mathbb{R}^{m \times n}$ satisfy the RIP of order $2 k$, and let $y=A x_S+\eta^{\prime}$, where $x_S$ is $k$-sparse.
Then the iterate generated by the pursuit step in Algorithm \ref{algorithm} when $b=1$ satisfies
\[
\|x^{(p+1)}-x_S\|_2 \leq \frac{\|\mathcal{H}_k(u^{(p)})-x_S\|_2}{\sqrt{1-\delta_{2 k}^2}}+\frac{\sqrt{1+\delta_k}}{1-\delta_{2 k}}\|\eta^{\prime}\|_2 .
\]
\end{lemma}

\begin{theorem}
Let $A \in \mathbb{R}^{m \times n}$ satisfy
\begin{equation} \label{thm2-ric}
\delta_{3 k}<\frac{(1+\gamma) \varphi^{-1}-(1-\gamma)}{2+(1+\gamma) \varphi^{-1}}.
\end{equation}
Choose $\lambda$ such that
\begin{equation} \label{thm2-lambda}
\frac{1-\varphi^{-1}\left(1-\delta_{3 k}\right)}{\gamma-\delta_{3 k}} < \lambda < \frac{1+\varphi^{-1}\left(1-\delta_{3 k}\right)}{1+\delta_{3 k}}.
\end{equation}
Then, for every $p \geq 0$, the iterates generated by SGHTP satisfy
\[
\|x^{(p+1)}-x_S\|_2 \leq \rho_2 \|x^{(p)}-x_S\|_2+\tau_2\|\eta^{\prime}\|_2,
\]
where 
\begin{equation} \label{thm2-tau}
\tau_2:=\frac{\varphi \lambda}{\sqrt{1-\delta_{2 k}}}+\frac{\sqrt{1+\delta_k}}{1-\delta_{2 k}} 
\end{equation}
and 
\[
\rho_2= \begin{cases} \varphi[1-\lambda(\gamma-\delta_{3 k})]/\sqrt{1-\delta_{2 k}^2}, & \textrm{ if } 0<\lambda \leq \frac{2}{1+\gamma} \\ \varphi[\lambda(1+\delta_{3 k})-1]/\sqrt{1-\delta_{2 k}^2}, & \textrm{ if }  \lambda>\frac{2}{1+\gamma} \end{cases}
\]
which is smaller than 1.

\end{theorem}

\begin{proof}
Each SGHTP iteration applies exact least-squares refinement after the hard-thresholding update.
From the analysis of SGHT, we have
\[
\|\mathcal{H}_k(u^{(p)})-x_S\|_2 \leq \varphi(\Delta_{\lambda, \gamma}+\lambda \delta_{3 k}) \|x^{(p)}-x_S\|_2+\tau_1 \|\eta^{\prime}\|_2.
\]
Combining this bound with Lemma \ref{lemma-pursuit} gives
\begin{equation} \label{thm2-1}
\|x^{(p+1)}-x_S\|_2 \leq \frac{\varphi\left(\Delta_{\lambda, \gamma}+\lambda \delta_{3 k}\right)}{\sqrt{1-\delta_{2 k}^2}} \|x^{(p)}-x_S\|_2+\tau_2\|\eta^{\prime}\|_2,
\end{equation}
where $\tau_2$ is given by \eqref{thm2-tau}.
It remains to show that the coefficient of $\|x^{(p)}-x_S\|_2$ in \eqref{thm2-1} is smaller than 1.
Since $\delta_{2 k} \leq \delta_{3 k}<1$, we have $1-\delta_{3 k} \leq \sqrt{1-\delta_{3 k}^2} \leq \sqrt{1-\delta_{2 k}^2}$.
Therefore, it is sufficient to require
$ 
\varphi\left(\Delta_{\lambda, \gamma}+\lambda \delta_{3 k}\right)<1-\delta_{3 k} .
$
We consider the two cases in the definition of $\Delta_{\lambda,\gamma}$.
For $0<\lambda \leq \frac{2}{1+\gamma}$, we have $\Delta_{\lambda, \gamma}=1-\gamma \lambda$, giving $\varphi\left[1-\lambda\left(\gamma-\delta_{3 k}\right)\right] < 1-\delta_{3k}$.
It implies
\begin{equation} \label{thm2-lambda1}
\frac{1-\varphi^{-1}\left(1-\delta_{3 k}\right)}{\gamma-\delta_{3 k}} < \lambda \le \frac{2}{1+\gamma}.
\end{equation}
When $\lambda > \frac{2}{1+\gamma}$, we have $\Delta_{\lambda, \gamma}=\lambda-1$, giving $\varphi\left[\lambda\left(1+\delta_{3 k}\right)-1\right] < 1-\delta_{3k}$
which implies
\begin{equation} \label{thm2-lambda2}
\frac{2}{1+\gamma}<\lambda<\frac{1+\varphi^{-1}\left(1-\delta_{3 k}\right)}{1+\delta_{3 k}} .
\end{equation}
Both intervals, \eqref{thm2-lambda1} and \eqref{thm2-lambda2}, are nonempty under the stated RIP condition in \eqref{thm2-ric}.
Their union gives the step-size range in \eqref{thm2-lambda}, as stated in the theorem.
Therefore, every $\lambda$ in this interval satisfies $\rho_2<1$, which completes the proof.
\end{proof}

\section{Numerical Experiments}
In the synthetic experiments, $A\in\mathbb{R}^{m\times n}$ has i.i.d.\ $\mathcal{N}(0,1/m)$ entries.
Each sparse signal $x^*$ has a uniformly random support and independent standard Gaussian nonzero entries.

\subsection{Effects of $q$ and $\gamma$ on recovery performance}
We examine how $q$ and $\gamma$ affect SGHT recovery with noisy measurements.
We set $n=1024$, $m=\lceil0.5n\rceil$, and $\lambda=1$, and vary $k$ over 20 equally spaced values from $\lceil0.02m\rceil$ to $\lceil0.35m\rceil$.
Gaussian noise is scaled to a measurement SNR of $30\,\mathrm{dB}$.
For each sparsity level, all parameter combinations are evaluated on the same 100 independent instances, with $q\in\{\lceil0.1k\rceil,\lceil0.2k\rceil,\ldots,\lceil0.9k\rceil\}$ and $\gamma\in\{0.1,0.15,\ldots,0.9\}$.
A trial is successful and terminates when $\mathrm{SNR}_{\mathrm{rec}}:=20\log_{10} (\|x^*\|_2/\|x^{(p)}-x^*\|_2)\geq30\,\mathrm{dB}$, or terminates after 30 iterations otherwise.

For each parameter setting, we fit the recovery success probabilities $P_{\mathrm{succ}}$ using $\operatorname{logit}(P_{\mathrm{succ}})=\beta_0+\beta_1\rho$, where $\beta_0,\beta_1$ are fitted coefficients and $\rho=k/m$.
The fitted $50\%$ recovery transition is $\rho_{0.5}=-\beta_0/\beta_1$, with larger values indicating recovery at higher sparsity ratios within the iteration budget.

\begin{figure}[h]
\centering
\includegraphics[width=0.75\columnwidth]{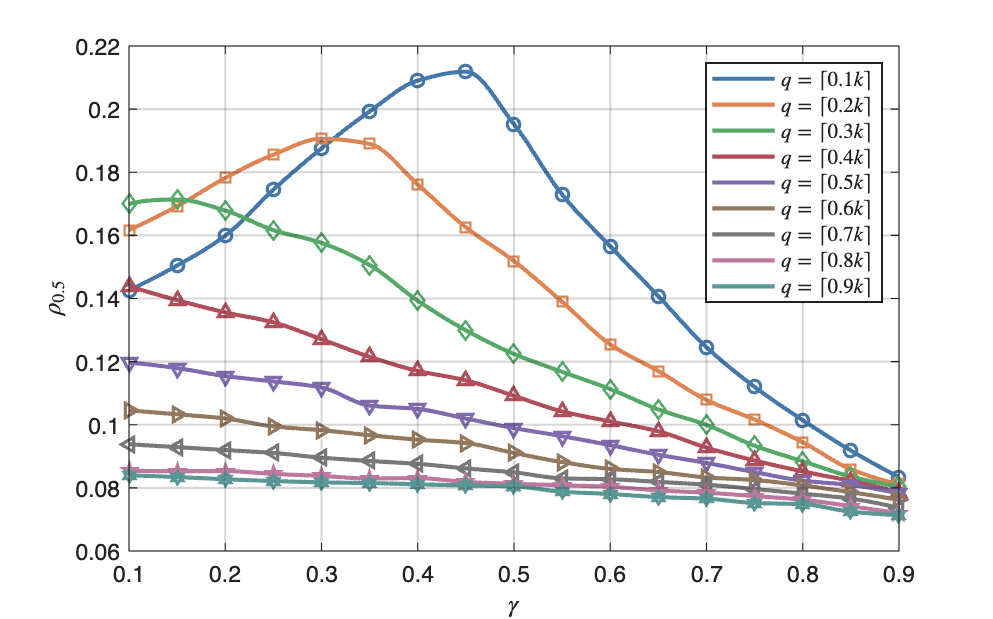}
\caption{Fitted $50\%$ recovery transition $\rho_{0.5}$ vs. $\gamma$ for different choices of $q$.}
\label{SGHT_rho05_gamma}
\end{figure}

Fig.~\ref{SGHT_rho05_gamma} shows that the effect of $\gamma$ on recovery performance depends on $q$.
For smaller $q$, the largest $\rho_{0.5}$ is attained at intermediate values of $\gamma$.
As $q$ increased, the best-performing values of $\gamma$ shift towards smaller values, and the sensitivity to $\gamma$ decreases.
Differences across $q$ also diminish as $\gamma$ increases.
These results also indicate that $q$ and $\gamma$ should be selected jointly.

\subsection{ECG Signal Recovery}
We compare SGHTP with HTP \cite{foucart2011hard}, subspace pursuit (SP) \cite{dai2009subspace}, CoSaMP \cite{needell2009cosamp}, and EDOMP \cite{zhao2023dynamic} on the MIT-BIH ECG Compression Test Database, available through PhysioNet \cite{moody1988evaluation,moody1999mit}.
We use the complete first channel of each of the 168 records, with $n=5120$.
Each signal is mean-centered and normalized to unit $\ell_2$ norm to obtain $s$.
A five-level sym4 wavelet decomposition with periodized boundary extension yields $x\in\mathbb{R}^n$.

\begin{figure}
\centering
\subfloat[ECG PRD recovery frequency.]{
\includegraphics[width=0.6\columnwidth]{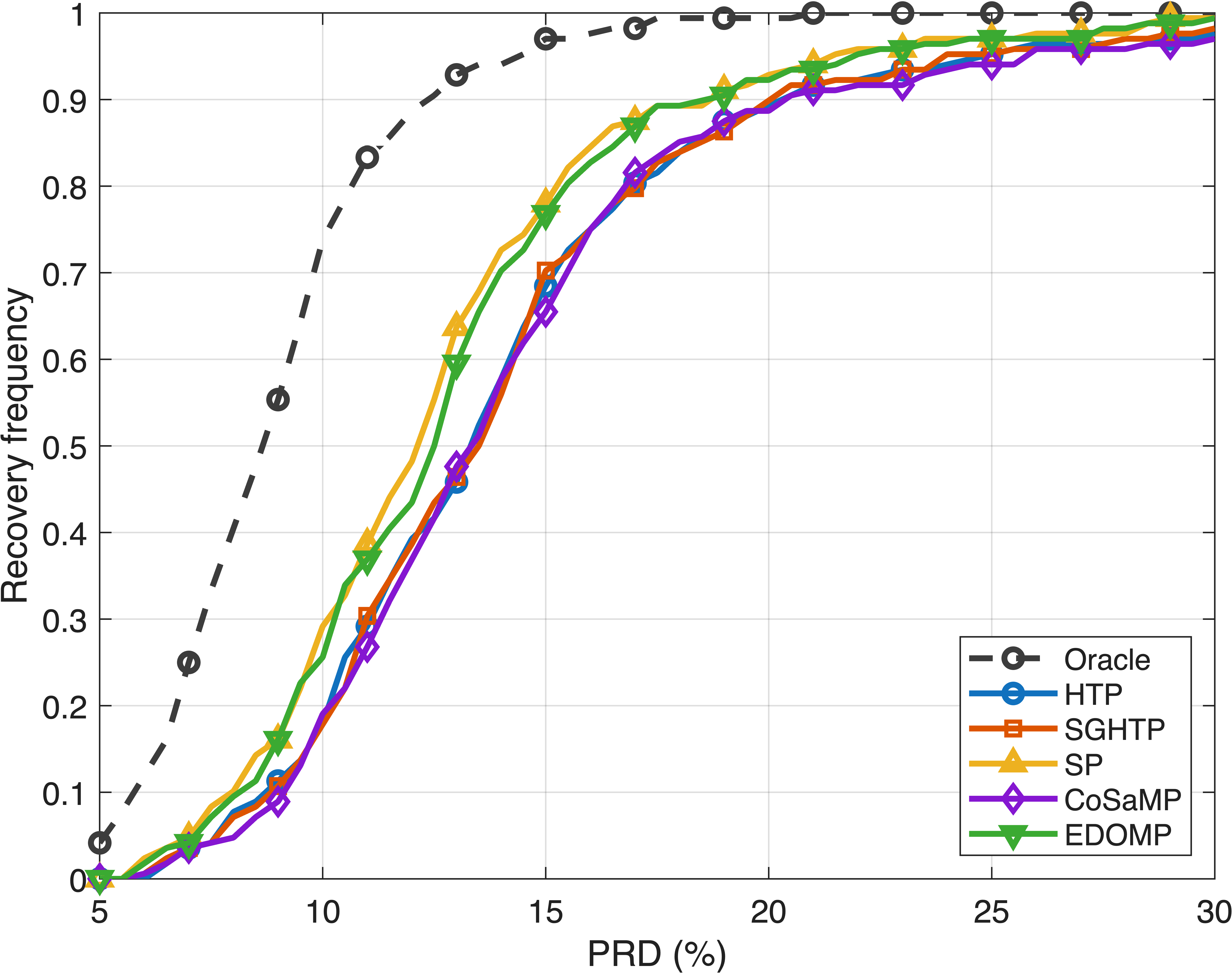}}

\subfloat[ECG average runtime under PRD thresholds.]{
\includegraphics[width=0.6\columnwidth]{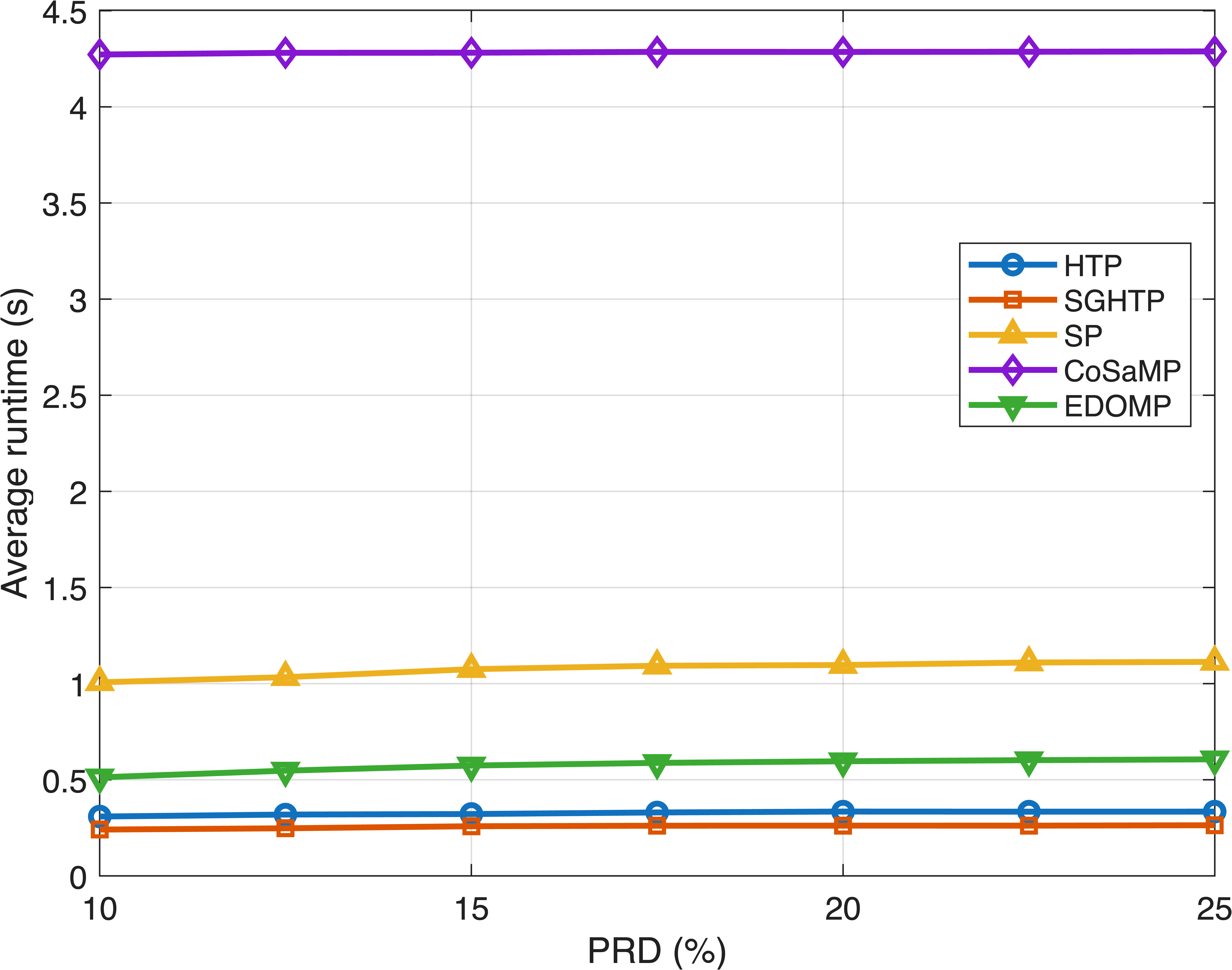}}

\caption{Recovery performance of different algorithms on the MIT-BIH ECG Compression Test Database.}
\label{ecg}
\end{figure}

Measurements are $y=Ax$, where $A\in\mathbb{R}^{m\times n}$ has i.i.d.\ $\mathcal{N}(0,1/m)$ entries and $m=0.4n$.
For each record, all methods use the same sensing matrix and measurements, with $k=0.1n$.
We set $\lambda=1$ for HTP and SGHTP, with $q=0.5k$ and $\gamma=0.3$ for SGHTP.
The gradient-selection parameter of EDOMP is $0.7$.
Each method terminates when the relative change between successive iterates is smaller than $10^{-6}$ or after 50 iterations, whichever occurs first.
The inverse wavelet transform of $\widehat{x}$ yields $\widehat{s}$.
Reconstruction quality is measured by the percentage root-mean-square difference, $\mathrm{PRD}=100\|\widehat{s}-s\|_2/\|s\|_2$.
The best $k$-term wavelet approximation provides the oracle PRD.
For each method, we report the fraction of records meeting each PRD threshold and the average runtime over those records.

Fig. \ref{ecg}(a) shows that SGHTP achieves recovery frequencies comparable to those of HTP across the tested PRD range and generally higher than those of CoSaMP. 
SP and EDOMP attain higher recovery frequencies at intermediate thresholds, although their advantage over SGHTP narrows as the threshold increases.
Fig. \ref{ecg}(b) shows that SGHTP has the lowest average runtime over records satisfying each displayed PRD threshold.
These results indicate that SGHTP achieves competitive reconstruction quality relative to the comparison methods on the tested ECG records, with low observed runtimes under the stated experimental settings.

\section{Summary}
We incorporated a component-wise gradient-scaling strategy into hard-thresholding methods for sparse optimization, yielding SGHT and SGHTP.
Under suitable RIP conditions, we established recovery error bounds for both methods in the presence of measurement noise.
Synthetic experiments highlight the importance of joint parameter selection, while experiments on the tested ECG records show recovery performance comparable to established methods.

\clearpage
\IEEEtriggeratref{15}

\section*{References}

\def\refname{}

\end{document}